\documentclass{aastex}          
\usepackage{spr-astr-addons}    
\usepackage{url}

\begin{document}

\title{Discovery of GeV gamma-ray emission from the LMC B0443-6657 with the \textit{Fermi} Large Area Telescope}

\shorttitle{LMC detection}
\shortauthors{Tang et al.}

\author{Qing-Wen Tang\altaffilmark{1,2}}

\email{qwtang@ncu.edu.cn} 

\altaffiltext{1}{Department of Physics, Nanchang University, Nanchang 330031, China, $^*$qwtang@ncu.edu.cn}
\altaffiltext{2}{Center for Cosmology and AstroParticle Physics (CCAPP),
Ohio State University, Columbus, OH 43210, USA}

\begin{abstract}
We report the discovery of gamma-ray detection from the Large Magellanic Cloud (LMC) B0443-6657 using the Large Area Telescope (LAT) on board the \textit{Fermi Gamma-ray Space Telescope}. LMC B0443-6657 is a flat spectrum radio source, possibly associated with a supernova remnant in the Large Magellanic Cloud (LMC N4). Employing the LAT data of 8 years, our results show a significant excess ($>9.4\sigma$) of gamma-ray in the range of 0.2--100\,GeV above the gamma-ray background. A power-law function is found to be adequate to describe the $0.2-100$~GeV $\gamma$-ray spectrum, which yields a photon flux of $3.27\pm0.53\  \mathrm{photon\ \,cm}^2\ \mathrm{s}^{-1}$ with a photon index of $2.35\pm0.11$, corresponding to an isotropic gamma-ray luminosity of $5.3\times10^{40}\ \mathrm{erg\ \,s}^{-1}$.
The hadronic model predicts a low X-ray and TeV flux while the leptonic model predicts an observable flux in these two energy bands. The follow-up observations of the LMC B0443-6657 in X-ray or TeV band would distinguish the radiation models of gamma-rays from this region.
\end{abstract}


\section{Introduction}\label{s:intr}
The Large Magellanic Cloud (LMC) is the brightest external galaxy in the
gamma-ray band, as it is very close to us (only about 49.97 kpc).
High-energy (HE, $>$100 MeV) diffusive gamma rays in the LMC,
which are thought to originate from the large-scale disk, are mainly produced by hadronic process~\citep{2015ApJ...808...44F,2016A&A...586A..71A,2017ApJ...843...42T}.
This disk component contributes a large amount of gamma rays of LMC observed by \textit{Fermi} Large Area
Telescope (LAT, 20~MeV to above 300 GeV). However, the details on the
distribution and origin of the CRs cannot be revealed
without detections of the individual sources in the LMC.
Recently, a few luminous point sources has been detected in the LMC, thanks to high-energy and very-high-energy (VHE, $>$100 GeV) telescopes. For example, two bright young pulsars (PSR J0540-6919 and PSR J0537-6910) are detected in gamma-ray band at the eastern of 30 Doradus in the LMC by Fermi-LAT~\citet{2015Sci...350..801F,2016A&A...586A..71A}.
VHE gamma-ray emission of PSR J0537-6910 is also detected with high significance by
the High Energy Stereoscopic System (H.E.S.S., \citealt{2015Sci...347..406H}).
The more HE/VHE detections of the individual LMC point sources would help us to better understand
the nature of the cosmic rays (CRs) in the LMC.

In this work, we report a new HE gamma-ray detection
in the northeast of the LMC employing the data of 8-yr LAT observations.
Searching the radio observations
in the vicinity of the new source, we find a spatial correlation between this gamma-ray point source
and a flat spectrum radio source, LMC B0443-6657.
Further analysis gives a constraint on the physical parameters of CRs and the
surrounding medium of the LMC B0443-6657.

\section{LAT data analysis}\label{s:data}

\subsection{Data selection and background templates}\label{ss:dataselect}

The 8-year gamma-ray data since 2008 August 4 are acquired in the \textit{Fermi} Science Support Center~({\url{https://fermi.gsfc.nasa.gov/ssc/}). Events with energy between 200~MeV and 100~GeV are selected using the instrument response function (IRSF) of
P8R2$\_$SOURCE$\_$V6. We exclude the events with zenith angles $>$90$^\circ$ to reduce the contribution of Earth-limb gamma rays. A box Region of interest (ROI) of 10$^\circ\times$10$^\circ$ centering at the position of RA=80.894$^\circ$, Decl.=-69.756$^\circ$ is selected in the Binned maximum likelihood~(\url{https://fermi.gsfc.nasa.gov/ssc/data/analysis/scitools/binned_likelihood_tutorial.html}).

Two background templates are constructed to model the background gamma-ray emission within the ROI. One of them is based on third Fermi-LAT source catalog (3FGL) sources, which allows us to search for the new sources from the background emission in the standard LAT source analysis. Another template is derived following the most recent results of spatial analysis (6.3 yrs observations) by Fermi-LAT team~\citep{2016A&A...586A..71A}, which provides the details on the associated-source distribution in the LMC.

First, we build a background template that includs the 3FGL sources (hereafter called ``T1''). It contains 9 point sources and 1 extended source (LMC, 3FGL0526.6-6825e). The flux amplitude and spectral index of these sources are left as free parameters in our fitting.

Second, removing 7 3FGL point sources in the H$\alpha$ region of LMC~\citep{2001PASP..113.1326G}, we construct another background template (hereafter called ``T2'') by adding 4 point sources (named as P1, P2, P3, P4) and 4 extended sources (named as LMCG1, LMCG2, LMCG3, LMCG4), whose positions and/or extended size are fixed as that derived by \textit{Fermi}-LAT Collaboration~\citep{2016A&A...586A..71A,2017ApJ...843...42T}. Note that, none of the 7 removed sources are associated with any known systems in the 3FGL with 4-yr observations.

For these two templates, a Galactic diffuse source and an isotropic gamma-ray source are added, which are represented by ``gll$\_$iem$\_$v06.fits'' and ``iso$\_$P8R2$\_$S\\OURCE$\_$V6$\_$v06.txt'' respectively.

\subsection{New gamma-ray source detection, spatial analysis and comparison of templates}\label{ss:method}

First, we make the spectral fits over the 0.2$-$100 GeV range
with 11 logarithmic energy bins and using 0.1$\times$ 0.1 pixels.
A common assumption of power-law photon spectrum is employed to describe the new sources.
The significance of a new source
is evaluated by the maximum significance value, i.e.,
\begin{equation}
\sigma_{\rm max}=\sqrt{ 2(\log L - \log L_0)}
\end{equation}
where $L$ and $L_0$ are the best-fit likelihood with and without a new source~\citep{1996ApJ...461..396M} respectively. A new source is detected significantly when $\sigma_{\rm max}$
is larger than 5 and labeled as S$\bf n$,
where $\bf n$ represents the $\bf n$-th significance source.

Secondly, we calculate the significance of spatial extension of each new source, defined as the change of the best-fit likelihood from a point-like source to an extended source:
\begin{equation}
{\rm \Delta TS_{ext}}=2(\ln L_{\rm ext}-\ln L_{\rm point})
\end{equation}
where $\ln L_{\rm point}$ and $\ln L_{\rm ext}$ are the best-fit likelihood values for the point-source model and the extended model. When the maximum ${\rm \Delta TS_{ext}}$ is larger than 25, the new source is considered as one extended source more than a point source.

Finally, results for T1 and T2 are compared based on their maximum
likelihood values ($\log L_{\rm TL}$) relative to the value that without LMC field sources($\log L_{\rm BR}$), which is defined as the changing TS (${\rm TS_{\rm TL}}$) value:
\begin{equation}
{\rm TS_{TL}}=2(\ln L_{\rm TL}-\ln L_{\rm BR})
\end{equation}
We assess a better template when one has fewer free parameters and a larger ${\rm TS_{TL}}$ value.

\subsection{Timing and spectral analysis}\label{ss:lcsed}

Timing analysis. For the new source, the light curve is calculated by dividing 8-year observations into 12 equal time bins. With all the parameters fixed except for the normalization of
the interesting source and two diffuse backgrounds in the
optimized model, the likelihood fits are performed independently in each time bin. Each resultant best-fit flux can be derived in the likelihood analysis.

Spectral analysis. To calculate the spectral energy distribution of a new source, we consider three possible photon models, (1) the Power Law model (PL):
\begin{equation}
{\rm dN/dE=N_0(E/100MeV)}^{-\Gamma_{\rm PL}}
\end{equation}
where ${\rm N_0}$ is the normalization and $\Gamma_{\rm PL}$ is the photon index,
(2) the Power Law with an exponential cutoff model (PLExpCut):
\begin{equation}
{\rm dN/dE=N_0(E/100MeV)}^{-\Gamma_{\rm PLE}}\exp(\rm -E/E_{\rm cut})
\end{equation}
where ${\rm N_0}$ is the normalization, $\Gamma_{\rm PLE}$ is the photon index and E$_{\rm cut}$ is the cutoff energy, and (3) the LogParabola model (LogPara):
\begin{equation}
{\rm dN/dE=N_0(E/E_b)}^{-\alpha+\beta\log(\rm E/E_b)}
\end{equation}
where ${\rm N_0}$ is the normalization, $E_b$ is the break energy, $\alpha$ is the photon index at the energy $E_b$ and $\beta$ measures the curvature of the parabola~\citep{2004A&A...413..489M}.

We adopt two methods to assess a good model.
One is to calculate the changing TS from a model to another model ($\Delta$TS).
For example, the PLExpCut is a nest model with the PL, thus
The PLExpCut model is better than the PL when $\Delta$TS$_{\rm PLExpCut \rightarrow PL}>$25~\citep{2013ApJS..209...11A}, i.e.,
\begin{equation}
{\rm \Delta TS}=2(\ln L_{\rm PLExpCut}-\ln L_{\rm PL})
\end{equation}
Another one is the Akaike information criterion (AIC)
test~\citep{1974ITAC...19..716A}.
The AIC is defined as:
\begin{equation}
{\rm AIC}=2k-2\log L
\end{equation}
where $k$ is the number of parameters of the corresponding
model and the model that has a smaller AIC is the
better one.

\section{Results}\label{s:results}

\subsection{Detection of a new point source in the LMC}\label{ss:detection}

Table \ref{tab:detection} and Figure \ref{fig:detection} show the results of the new gamma-ray detections in two background templates of T1 and T2. Four new sources are found in T1 while only one source is found in T2. A new source, S1, is the only source detected significantly above 200~MeV in both templates.
The locations of S1 are reported in Table \ref{tab:detection}. As seen, the detection significance of S1 is 11.1$\sigma$ in T1 and 9.4$\sigma$ in T2.

For S1, we do not find evidence for spatial extension. The maximum ${\rm \Delta TS_{ext}}$ is 2.8 in T1 and 2.0 in T2 respectively, both of which are less than 5$\sigma$ (${\rm \Delta TS_{ext}=25}$). S1 is eventually considered as a point source.

Comparing the results in the two templates, T2 is a better template for modeling the LMC ROI photons than T1 based on two facts: (1) TS$_{\rm TL}$ in T2 is larger than that in T1, i.e., 12064.8 in T1 and 12227.0 in T2, (2) the distribution of significance residuals in T2 is more satisfactory than that in T1, say, T2's has a mean closer to zero with a well balance between the negative and positive residuals. We thus focus on the analyses with the template T2 in the following.

We also note that the detection of S1 can be reproduced with other event selections, such as employing the photons above 1~GeV or centering the ROI in the position of S1, strongly suggesting that S1 is a real gamma-ray source.

\subsection{Light curve and spectral energy distribution of S1}\label{ss:resultlcsed}

The light curve of S1 is shown in Figure \ref{fig:lcsed}, with the average energy flux of $2.24\pm0.32\ {\rm MeV\ cm^{-2}\ s^{-1}}$ (grey horizon region), which is calculated in the best fit of 8-year observations. For the first 4 bins, the detection significances of S1 are low in the energy of 0.2-100~GeV, with TS of 0.26, 9.81, 1.02, 4.62 respectively, which is consistent with the non-detection of S1 in the 4-yrs source catalog. To evaluate the time variation, we measure the goodness of fit of the light curve excluding the two upper limits. The result shows a $\chi^2$ of 9.1 with 10 degrees of freedom (DOF), which implies a lack of temporal variability.

The spectral energy distribution of S1 is shown in Figure \ref{fig:lcsed}.
For three candidate photon models, the LogPara model ($k_{\rm LogPara}$=31) has two more free parameters than that in the PL model ($k_{\rm PL}$=29) and one more than that in the PLExpCut model ($k_{\rm PLExpCut}$=30). Their likelihood values ($L$) are -20932, -20929 and -20929 for the PL, LogPara and PlExpCut, respectively, which implies that both LogPara and PLExpCut model are not improving the fitting goodness to the PL significantly, i.e., $\Delta$TS$=6<$25. In the method of the Akaike information criterion (AIC) test, the PLExpCut gives the smallest AIC value (AIC$_{\rm PLExpCut}$=41018) than others, i.e., AIC$_{\rm PL}$=41022 and AIC$_{\rm LogPara}$=41020, however, the differences are not significant. We thus conclude that the PL model is still a reference model for simplicity.

\subsection{Associations of S1 in other wavelengths}\label{ss:associatioins}

We search for associations of S1
in four catalogs, i.e., the CRATES catalog of flat-spectrum radio sources~\cite{2007ApJS..171...61H}, the Veron-Cetty Catalog of Quasars \& AGN~\cite{2010A&A...518A..10V},
the Candidate Gamma-Ray Blazar Survey catalog (CGRaBS)~\cite{2008ApJS..175...97H} and the ATNF Pulsar Catalog version~\cite{2005AJ....129.1993M}.
Only one source, CRATES J044318-665155, is found 2.2$^\prime$ away from S1, which is within the 95\% C.L. localization radii of S1 (5.4$^\prime$) (see Table~.\ref{tab:detection}).

CRATES J044318-665155 is a flat spectrum radio sources and associated with the LMC B0443-6657, which has been observed by the Parkes radio telescope (FHW95,~\citealt{1995A&AS..111..311F}), the Parkes-MIT-NRAO Survey (PMN J0443-6651,~\citealt{2003MNRAS.342.1117M}) and the Sydney University Molonglo Sky Survey (SUMSS J044318-665204,~\citealt{1994ApJS...91..111W}). The radio observations from these detections are summarized in Table \ref{tab:radio} and Figure \ref{fig:radio}.
LMC B0443-6657 is thus considered as the counterpart of S1 in the radio band.

\section{Discussion}\label{s:discussion}

\subsection{Physical modeling with a typical magnetic field intensity of LMC}\label{ss:cosmic}

Recently, fast rises of gamma-ray fluxes below 1\,GeV and spectral breaks in the gamma-ray emissions of two Galactic supernova remnants (extended emission from IC 443 and W44,~\citealt{2013Sci...339..807A}) and LMC large-scale disk~\citep{2017ApJ...843...42T} have been discovered around 1\,GeV, which favor a hadronic origin rather than a leptonic one.
Such features, however, do not clearly appear in the discovered source in this work, and hence the leptonic model should also be discussed. In the following part of this section, we model the broadband emission with a one-zone model, say, all injected CR electrons or protons follow one simple power-law distribution.

We firstly model the broadband emission of the LMC B0443-6657 with the leptonic model.
Assuming a power-law spectrum for the injected CR electrons, we calculate their bremsstrahlung radiation, inverse-\textit{Compton} radation and synchrotron radiation~\citep{2006PhRvD..74c4018K,2015ApJ...808...44F,2017ApJ...843...42T}.
The result can be found in Figure \ref{fig:model}.
In the gamma-ray band, the relativistic bremsstrahlung radiation could contribute the most flux to the LAT observation. In the radio band, the high flux cannot be fitted by the synchrotron radiation of a single electron component which produce the GeV gamma rays, thus a second component arising from thermal bremsstrahlung radiation of thermal electrons is added to explain the data.
The best-fit result returns a PL index ($\Gamma_e$) of -1.82 for the electron spectral distribution, a magnetic field intensity ($B$) of ${\rm 5.35 \ \mu G}$, a photon energy density (${\rm U_{ph}}$) of $0.04 {\rm\ eV cm^{-3}}$ and a hydrogen atom density ($ n_{\rm H}$) of ${\rm 0.16 \ cm^{-3}}$.
The temperature of the thermal plasma which are responsible for the radio emission is around 1~eV (converted from the peak energy), but their origin is not clear. Indeed, given physics underlying the source itself is not clear, it is difficult to give an convincing explanation of such a component. Nevertheless, they could be an implication of existence of large amount of warm plasma around or in the source. Alternatively, as modeled in Sec. \ref{ss:prospect} with a large $B$, the synchrotron component can dominate the radio emission of LMC B0443-6657, in which case such a thermal component is not needed.

We also use a hadronic model to reproduce the broadband emission. The gamma-ray emission from neutral pion decay is calculated with the same method depicted in~\citet{2017ApJ...843...42T}. The result can be found in Figure \ref{fig:model}. The best-fit proton index is $\Gamma_p=$2.53. In this case, the CR electrons play a subordinate role, and the synchrotron radiation of CR electrons should be limited (lower than that in case of the leptonic model) in order not to overproduce the gamma-ray flux by their inverse Compton radiation. Thus, a similar thermal bremsstrahlung emission would be needed to explain the emission in the radio energy bands.

\subsection{Revised model with a large $B$ and prospect in X-ray and TeV bands}\label{ss:prospect}

Inverse-\textit{Compton} and Bremsstrahlung emission of electrons contribute about 92\% gamma-ray flux of 30 Doradus ~\citep{2015ApJ...808...44F}.
The current physical model of the LMC B0443-6657 is similar
to the case of the 30 Doradus, which motivate us to
consider the possibility of the existence of numerous relativistic electrons around the LMC B0443-6657.
When these relativistic electrons undergo synchrotron process or inverse-\textit{Compton} process with a large $B$, LMC B0443-6657 is expected to be detected in the X-ray or TeV band, at a flux level of about $10^{-14} \mathrm{erg\ cm^2 \,s}^{-1}$ as seen in Figure \ref{fig:pred}.

Effort are made to search the X-ray or TeV emission around the
LMC B0443-6657, however, we do not find the significant detection in both energy bands with current observations.
The follow-up observations of the LMC B0443-6657 in X-ray or
TeV band would help constrain the contribution from the
CR electrons and protons. Chandra X-ray Observatory and long-term monitoring of CTA are sensitive enough to detect X-ray and TeV emission with the predicted flux level. If these emissions are detected in the future, the physical origin of emission from the
LMC B0443-6657 can be regarded as the leptonic processes, say, CR electrons play the important role to produce the broadband emission.

\begin{table*}[ht]
\small
\caption{Gamma-Ray detection of S1 in two LMC source templates\label{tab:detection}}
\begin{tabular}{@{}crrrrrrrrrrr@{}}
\tableline
Template	&	\multicolumn{1}{c}{$\log L_{\rm T}$\tablenotemark{a}} 	&	 \multicolumn{1}{c}{TS$_{\rm TL}$\tablenotemark{b}} &$N_{\rm dof}\tablenotemark{c} $&	 \multicolumn{1}{c}{$\sigma_{\rm max}$\tablenotemark{d}} 	 &	 ${\rm RA}$\tablenotemark{d} 	&	 ${\rm Decl._ {S1}}$\tablenotemark{d}   &	 $r_{68}$\tablenotemark{d} 	& $r_{95}$\tablenotemark{d} 		 \\
&	&	 &	&	&Deg.		&Deg.		&Deg.	&	Deg.		 \\	
\tableline
Background	&	-27045.3 	&	...  &	9	&	... 	&	... 	&	... &	 ... &	...\\	
T1	&	-21012.9 	&	12064.8  &	32	&	11.1 	&	70.901 	&	-66.846 &	 0.051 &	0.083\\	 T2	&	-20931.8 	&	12227.0  &	29	&	9.4 	&	70.857 	&	-66.833 &	 0.056 &	0.090\\	 \tableline
\end{tabular}
\tablenotetext{a}{The best-fit likelihood value}
\tablenotetext{b}{The changing TS value from one template from the background}
\tablenotetext{c}{The number of the free parameter}
\tablenotetext{d}{Detection significance, localization and the corresponding error (68\%, 95\% confidence level)}
\end{table*}

\begin{table*}[ht]
\small
\caption{Radio observations in the vicinity of S1\label{tab:radio}}
\begin{tabular}{@{}ccccccccc@{}}
\tableline
Telescope or Survey	&	Frequency &    Name         &	RA\tablenotemark{a}  & Decl.\tablenotemark{a} &	Beam\tablenotemark{a}   & Flux\tablenotemark{a}		 &\multicolumn{1}{c}{Ref.\tablenotemark{b}}\\	
		&	MHz    &                 &  Deg. &  Deg.  &	arcsec   & mJy   &  \\		
\tableline
Parkes 	&	1400  & LMC B0443-6657 (FHW95 a)   & 70.927 &	-66.854    &	 912  &	155$\pm$31	 & (1) \\	
Parkes 	&	2450  & LMC B0443-6657 (FHW95 b)   & 70.814 &	-66.858    &	 531  &	146$\pm$29	&... \\	
Parkes 	&	4750  & LMC B0443-6657 (FHW95 c)   & 70.851 &	-66.864    &	 288  &	168$\pm$34	&... \\	
Parkes 	&	8550  & LMC B0443-6657 (FHW95 d)   & 70.869 &	-66.876    &	 162  &	136$\pm$26	&... \\	
PMN 	&	4850   & PMN J0443-6651  & 70.825 &	-66.865    &	 168  &	123$\pm$25	 & (2)	 \\
SUMSS 	&	843 & SUMSS J044318-665204  & 70.825 &	-66.868    & 18  &	125$\pm$25	 & (3) \\ \tableline
\end{tabular}
\tablenotetext{a}{The position, error and flux of the radio observation}
\tablenotetext{b}{(1)\cite{1995A&AS..111..311F}; (2)\cite{2003MNRAS.342.1117M}; (3)\cite{1994ApJS...91..111W}}
\end{table*}

\begin{figure*}[ht]
\centering
\includegraphics[width=2\columnwidth]{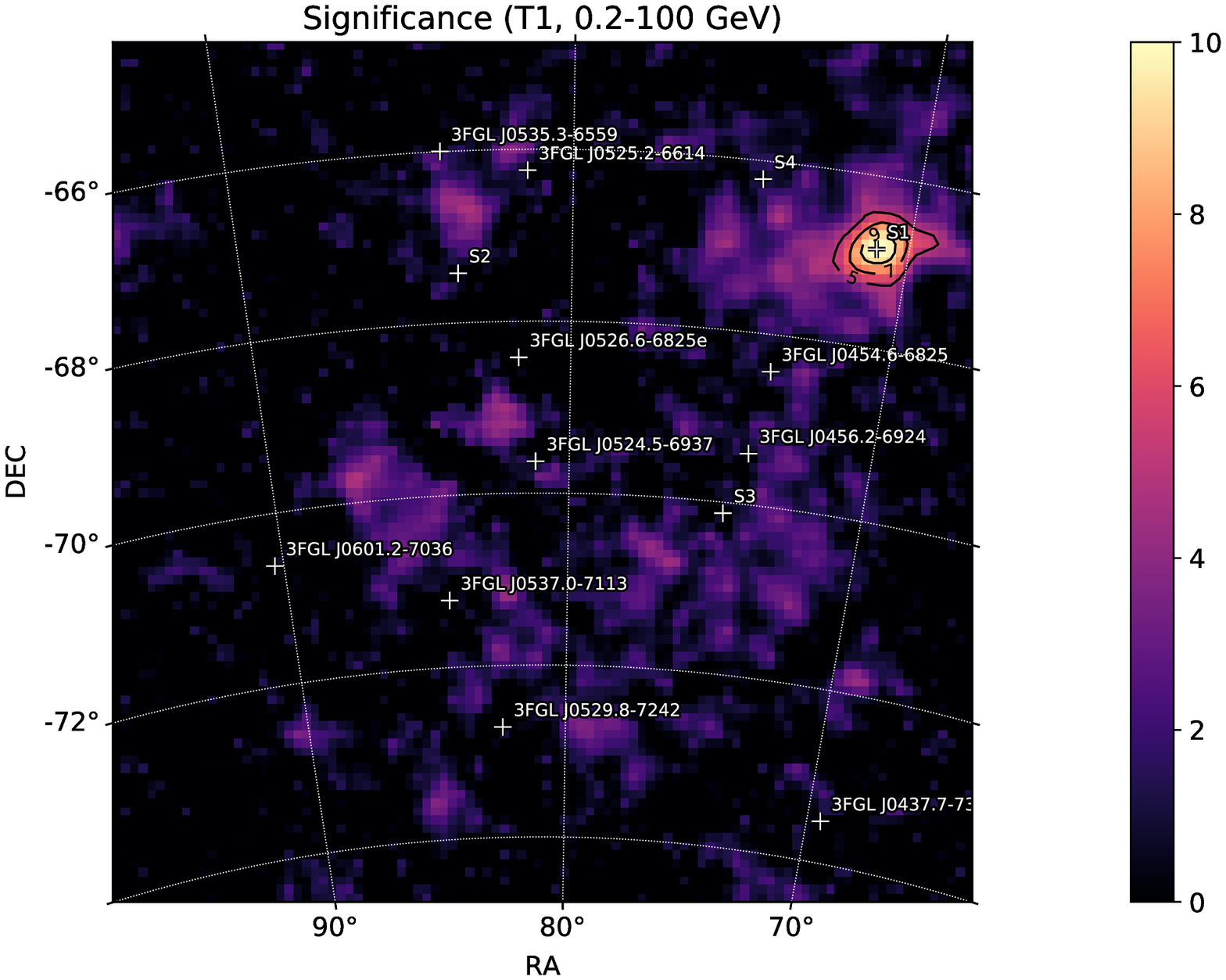}
\includegraphics[width=2\columnwidth]{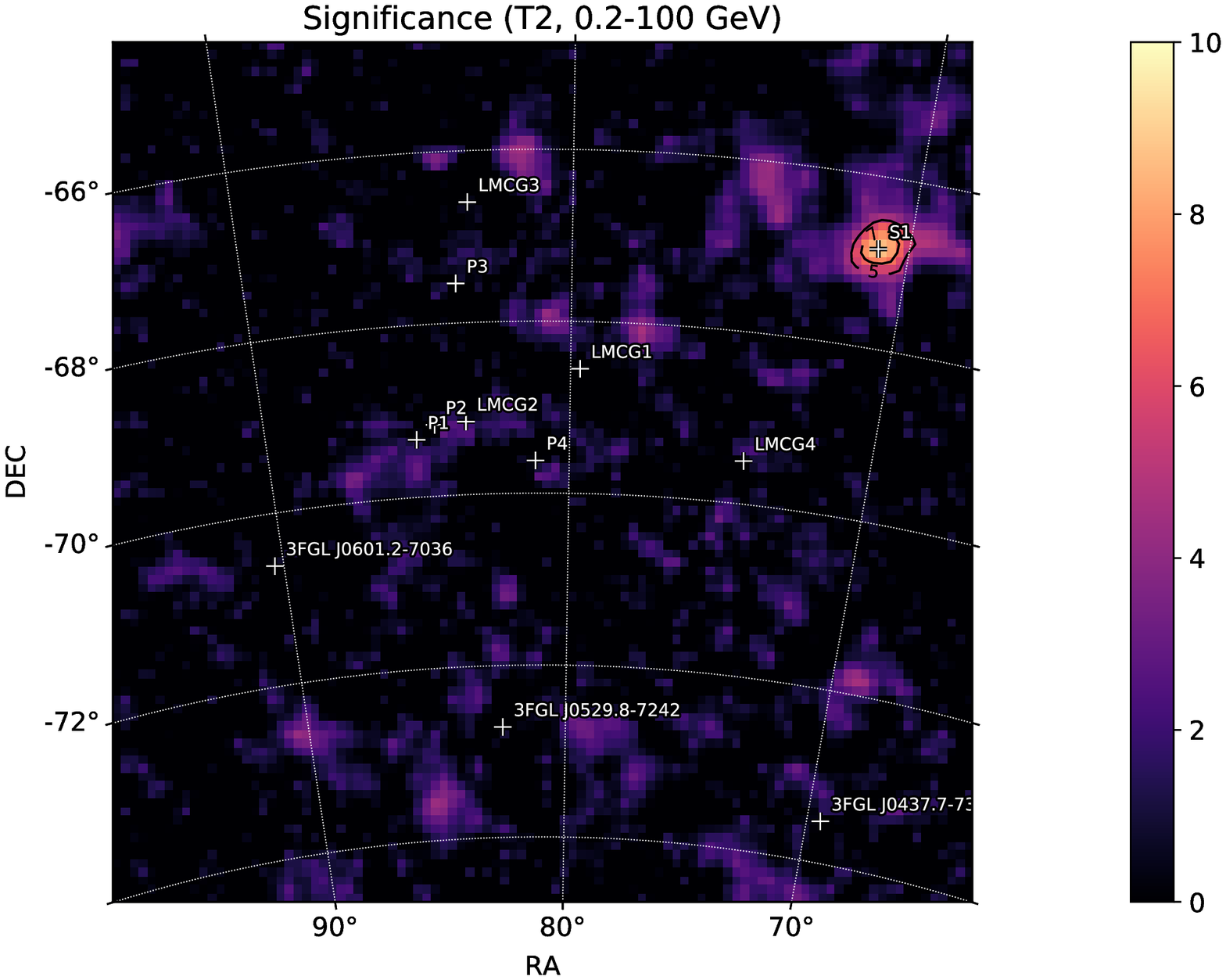}
\caption{Significance maps with photon energy between 0.2-100~GeV observed by \textit{Fermi}-LAT, top for T1 and bottom for T2. The maximum detection significance for the S1 are 11.1$\sigma$, 9.4$\sigma$ for T1 and T2 respectively.}
\label{fig:detection}
\end{figure*}

\begin{figure*}[ht]
\includegraphics[width=2\columnwidth]{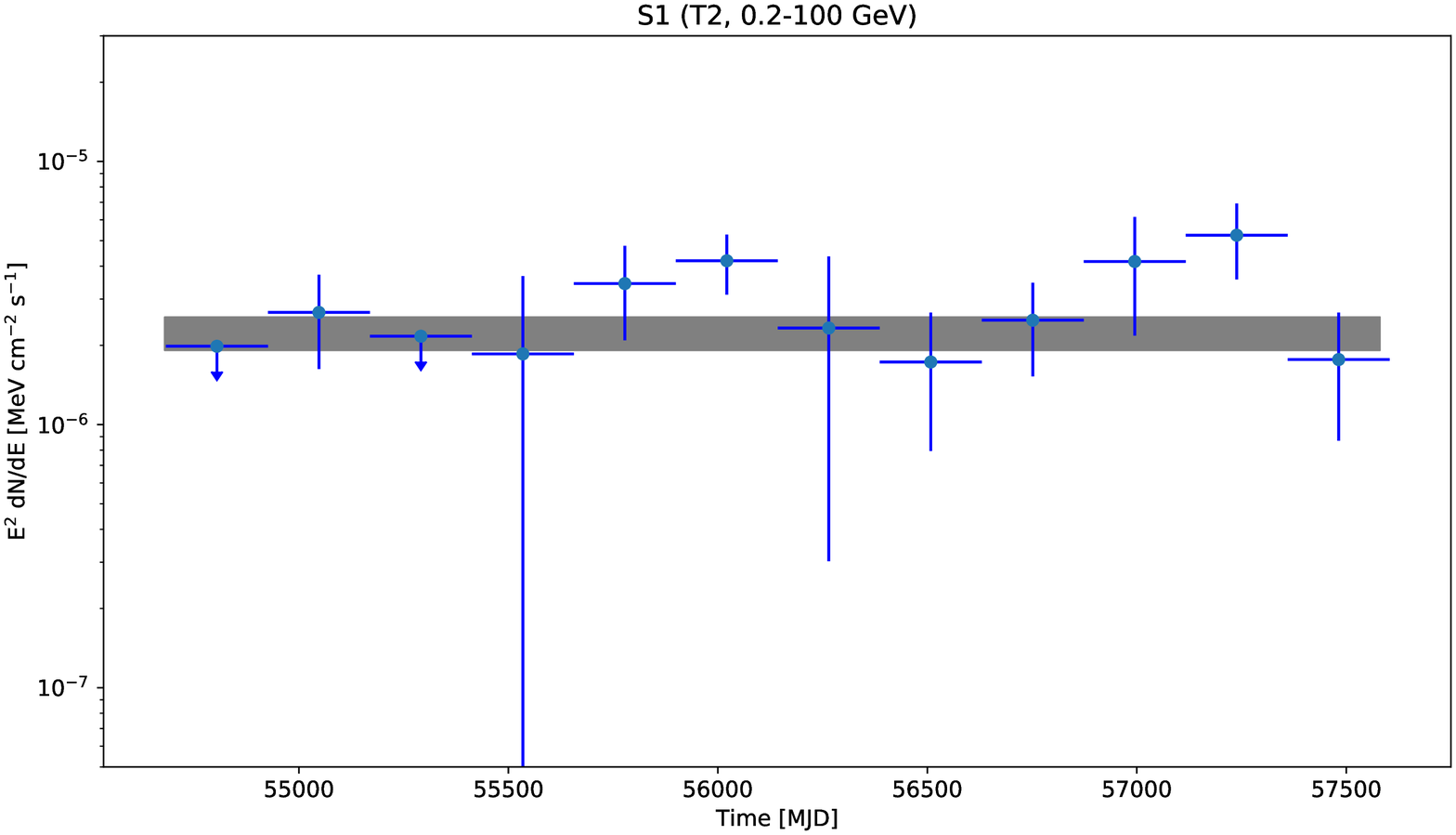}
\includegraphics[width=2\columnwidth]{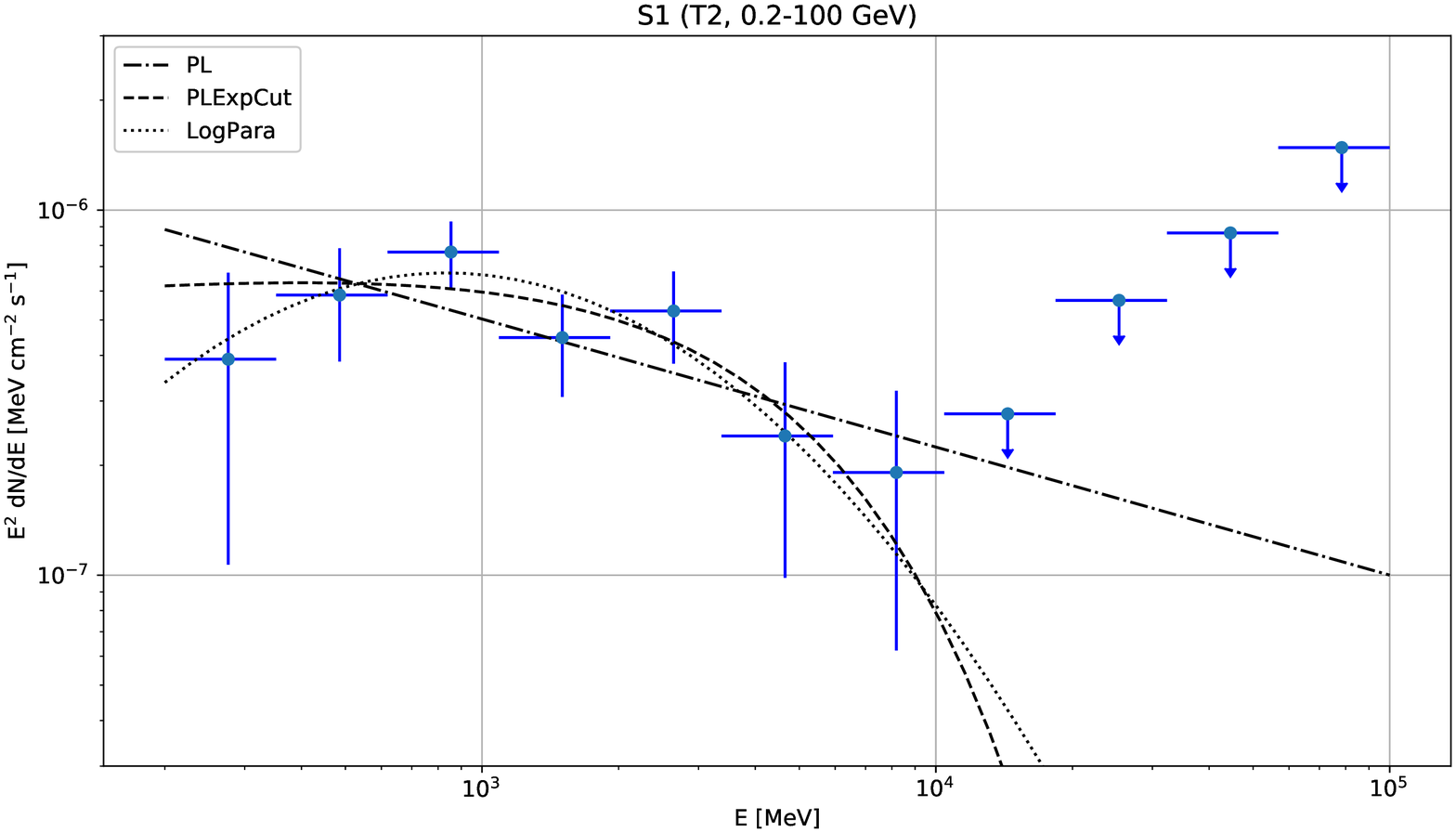}
\centering
\caption{Light curve and spectral energy distribution of S1 measured by the \textit{Fermi}-LAT. \textbf{Top:} Time since 2008 August 4, each bin is 243.5 days. The shaded region represents the best-fit average flux of 8-year observations. \textbf{Bottom:} The dot-dashed line is the PL model, the dashed line is the PLExpCut model and the dashed line is the LogPara model. All upper limits are at confidence level of 95\% .}
\label{fig:lcsed}
\end{figure*}

\begin{figure*}[ht]
\centering
\includegraphics[width=1.6\columnwidth]{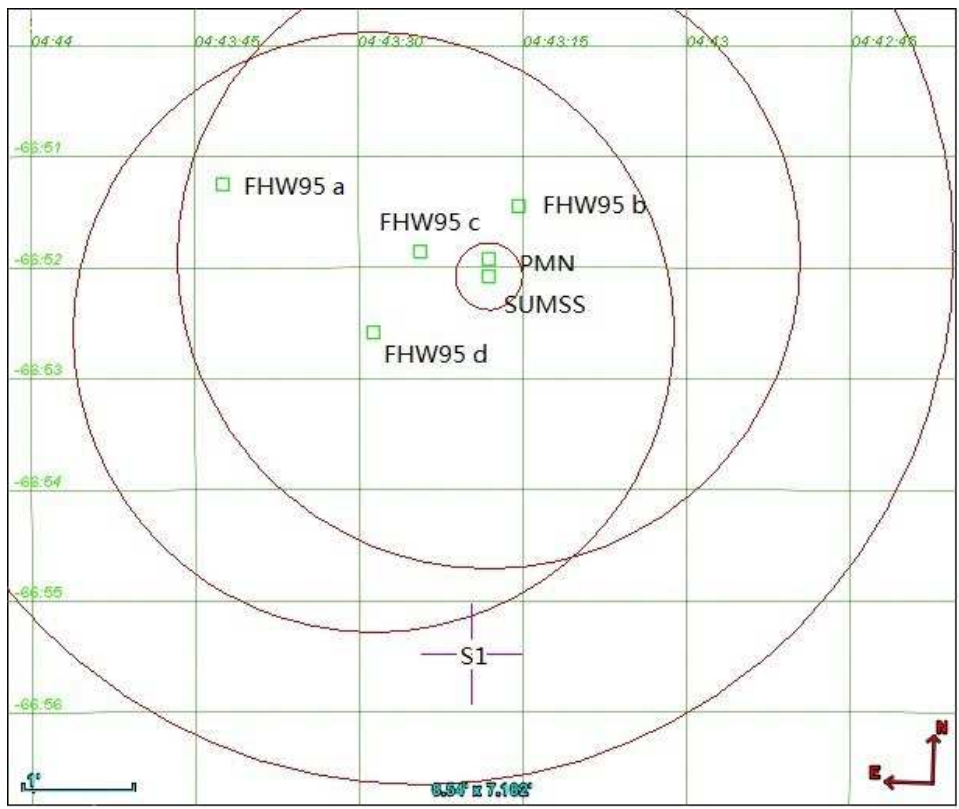}
\caption{Radio observations in the vicinity of of S1. Red cross is the center of S1 in gamma-ray band, green squares are the center positions of radio observations with corresponding beam (red circle). The size of per small box is $0.25^\prime\times1^\prime$.}
\label{fig:radio}
\end{figure*}

\begin{figure*}[ht]
\centering
\includegraphics[width=1.6\columnwidth]{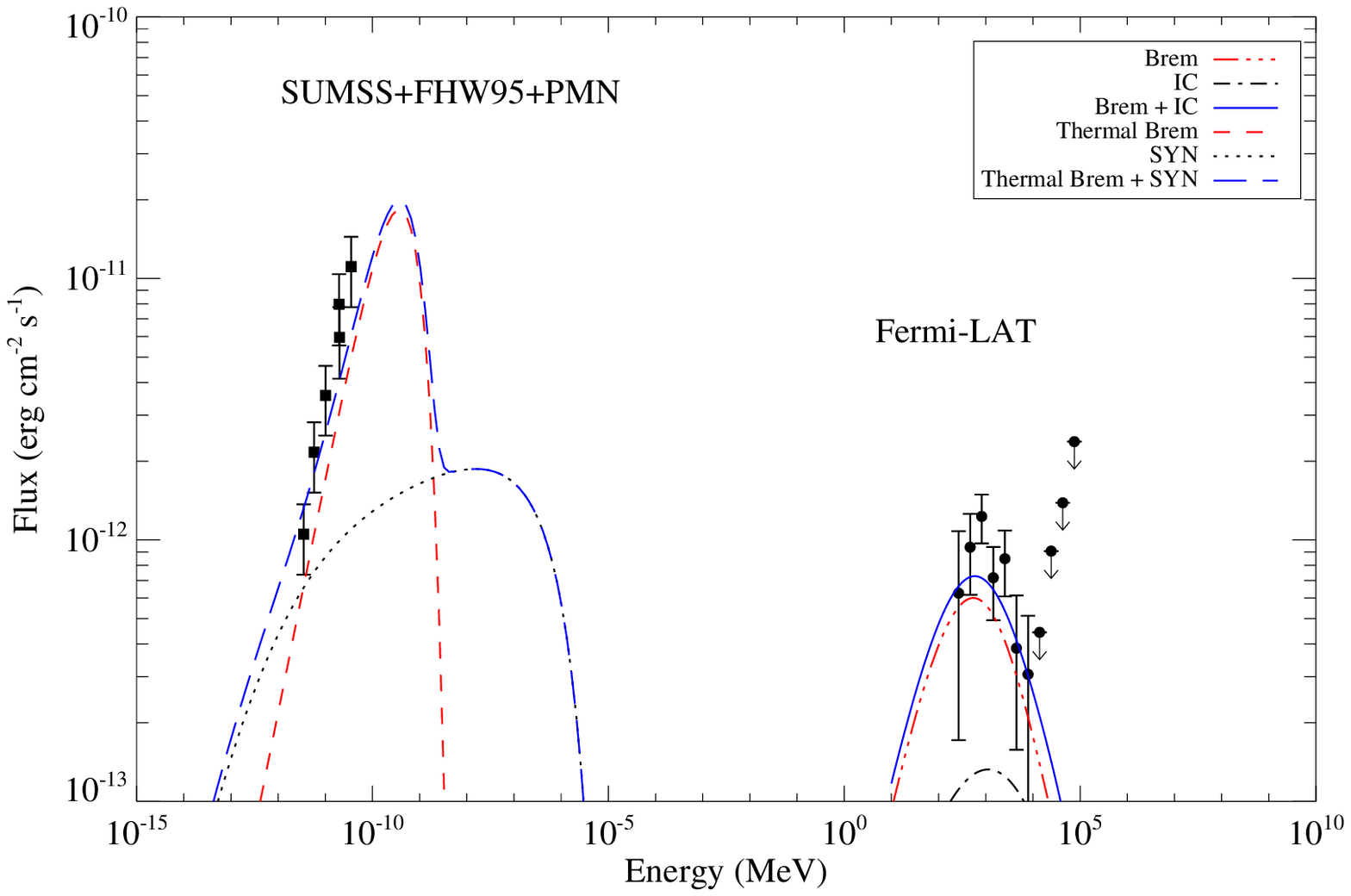}
\includegraphics[width=1.6\columnwidth]{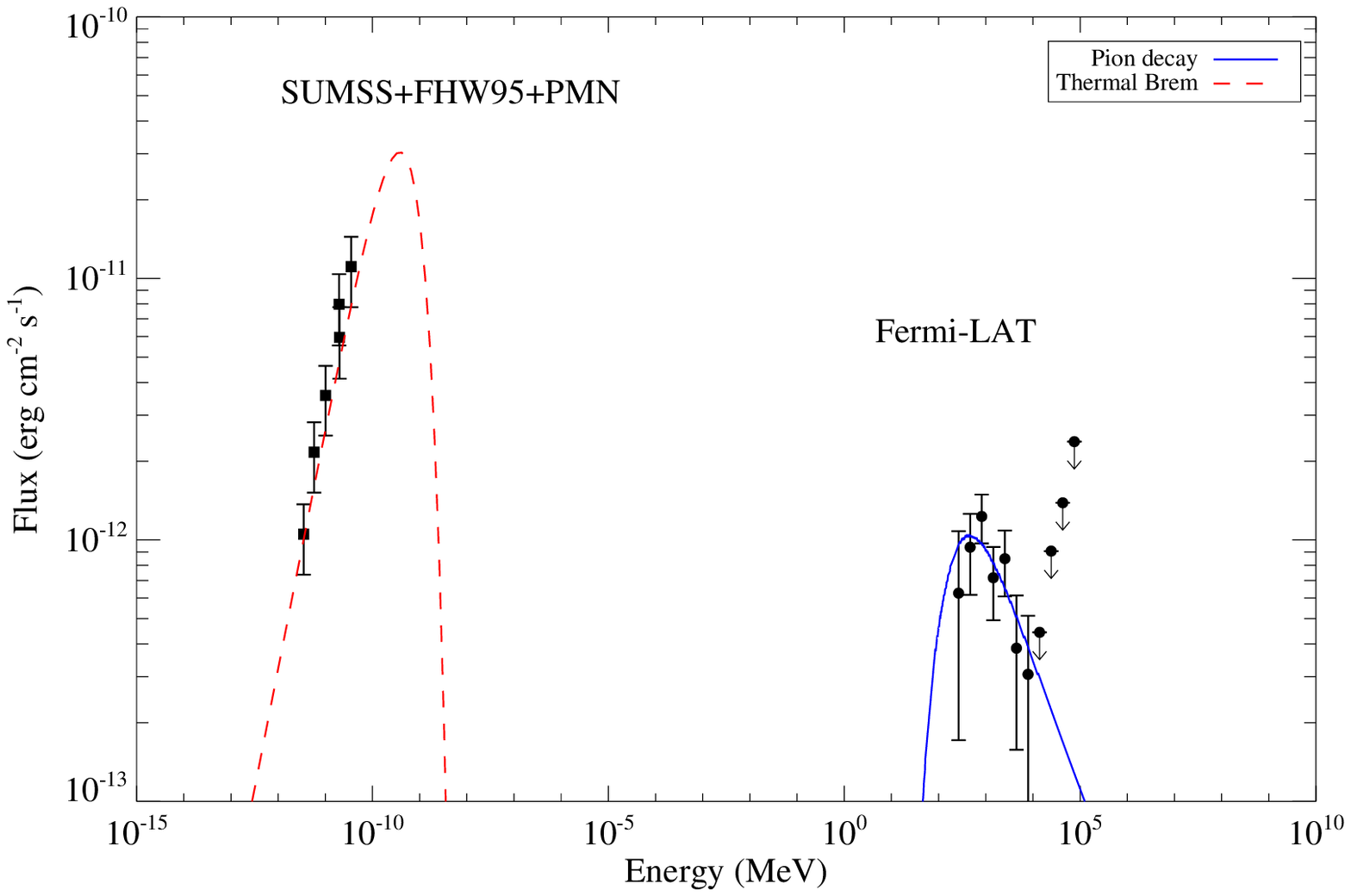}
\caption{\textbf{Top:} Leptonic modeling result in the broadband spectral energy distribution of the LMC B0443-6657. Here, $n_{\rm H} =0.16 {\rm \ cm^{-3}}$, $B = 5.35{\rm \ \mu G}$, U$_{\rm ph} = 0.04 {\rm\ eV cm^{-3}}$ and $\Gamma_e=-1.82$. \textbf{Bottom:} A pion decay model for the HE gamma-ray emission with a proton spectral index ($\Gamma_p$) of 2.53. The thermal component is derived with the same magnetic field strength ($B$) same as that in the top panel, but with a different normalization when fitting, i.e., about 1.33 times of that in the top panel.
The radio observations are carried out by the telescope or survey of Parkes, PMN, SUMSS as in Table \ref{tab:radio}. The gamma-ray data are derived from the \textit{Fermi}-LAT.}
\label{fig:model}
\end{figure*}

\begin{figure*}[ht]
\centering
\includegraphics[width=1.6\columnwidth]{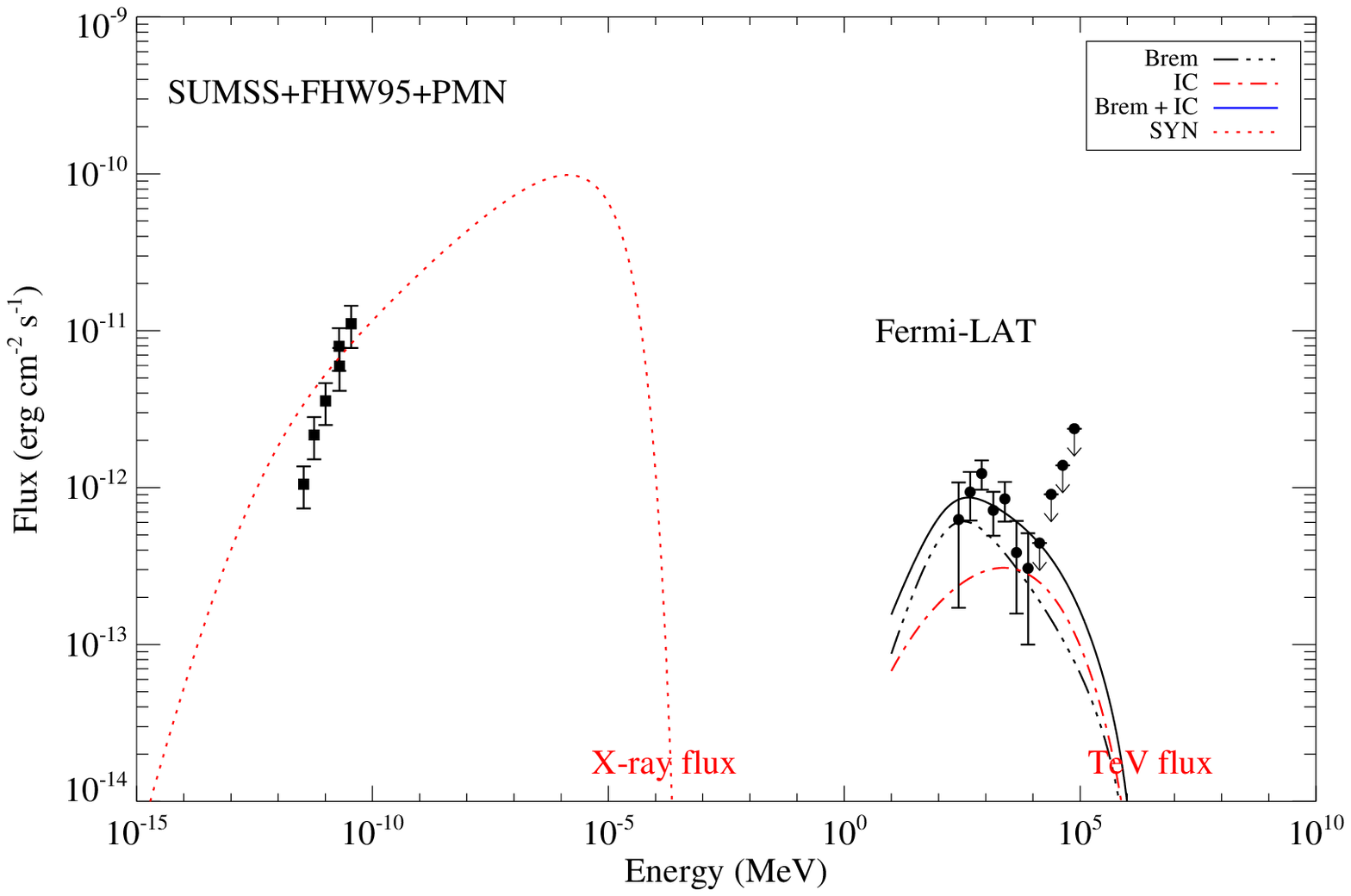}
\caption{Same as Figure \ref{fig:model} but with a large magnetic field intensity, i.e., $B = 100{\rm \ \mu G}$.
The X-ray and TeV flux are predicted in the leptonic model.}
\label{fig:pred}
\end{figure*}

%
\acknowledgments
I thank the anonymous referee for constructive comments and the editor for the helpful suggestion.
This work has made use of the programs written by Ruo-Yu Liu.
I am specially grateful to Ruo-Yu Liu,  Kwan-Lok Li and Francesco Capozzi for the final revision, and Xiang-Yu Wang for the helpful suggestion.
This work is supported by the Natural Science Foundation of China
under grants No. 11547029, 11533004, the Youth Foundation
of Jiangxi Province (No. 20161BAB211007), the Postdoctoral Foundation of Jiangxi Province (No. 2016KY17), the Natural Science Foundation of Jiangxi Provincial Department of Education (No. GJJ150077).

\end{document}